\documentclass[10pt,letterpaper]{article}
\usepackage{opex3}
\usepackage{amsmath,amssymb,cite,hyperref}
\begin{document}
\title{Time-resolved imaging with OKE-based time-gate: enhancement in spatial resolution using low-coherence ultra-short illumination}
\author{Harsh Purwar,$^*$ Sa\"id Idlahcen, Claude Roz\'e, and Jean-Bernard Blaisot}
\address{CORIA UMR-6614, Normandie Universit\'e, CNRS, Universit\'e et INSA de Rouen, \\Avenue de l'Universit\'e, 76800 Saint Etienne du Rouvray, France}
\email{$^*$harsh.purwar@coria.fr}
\begin{abstract}
We propose a collinear optical Kerr effect (OKE) based time-gate configuration with low coherence illumination source, derived from the supercontinuum (SC) generated by focusing the femtosecond laser pulses inside water. At first the spectral broadening in SC generation and corresponding changes in its coherence properties are studied and then a narrow band of wavelengths is extracted to use as the probe beam in the OKE-based time-gate configuration. The gate timings and spatial resolution of the time-gated images are also investigated. The low coherence of the probe ensures that the artifacts due to speckles from the laser are reduced to a minimum. To illustrate this a comparison of the time-resolved images of the fuel sprays obtained with this configuration has been made with the images obtained with the collinear, dual color configuration of the optical gate with coherent illumination.
\end{abstract}


\section{Introduction}
\label{sec:intro}
Over the past few years, due to technological advancements in the laser light sources and detection devices, there has been numerous efforts in developing optical diagnostic tools for applications in various fields of science and technology, since most of these tools are non-intrusive in nature. However, many real-world applications and phenomena of interest are intrinsically linked to turbid environments where light scattering and attenuation strongly limit the interpretation of optical signals. In a wide range of applications, from imaging in biological tissues \cite{Alfano1997,Wang1991a}, to measurements of high-pressure multiphase flows involving cavitation or turbulent breakup \cite{Linne2013a}, key information is scrambled by the distortion imparted to the light signal as it transits the measurement volume \cite{Berrocal2007}. Informative optical diagnostics in such media require detailed understanding of the light source, propagation and scattering in the measurement volume, and a detection arrangement tailored to collect the meaningful parts of the transmitted light signal. Assuming a vanishingly short input pulse and a $\sim1$~cm turbid measurement volume, the typical transmitted pulse durations can be expected to be of the order of $50-100$~ps, with most of the informative signal present in the first few picoseconds of the transmitted signal \cite{Berrocal2009}. Consequently, most time gating applications require arrangements that can limit light collection to a few picoseconds or shorter time window.

To this end, ultrafast time gating can provide an effective means of segregating high integrity portions of the collected signal from light disturbed by scattering interactions. On an average, photons that participate in more interaction events traverse a more circuitous path through the medium and hence their trajectories are statistically more likely to be distorted or redirected from their original trajectories. This difference in the optical path length results in a temporal spreading of the light. Time gating, or time filtering allows selection of the signal that retains the information on the object characteristics especially in case of high-pressure multiphase flows.

However, a common problem experienced by researchers trying to develop an ultrafast time-gated imaging technique for ballistic imaging or time-resolved imaging in general \cite{Duran2015,Rahm2014,Idlahcen2011,Linne2005,Wang1991}, is due to the illumination source itself, where an ultra-short laser pulse is used as imaging or probe pulse to illuminate the object under study. The unavoidable artifacts arising due to the speckles, produced by the laser source itself could be removed if the coherence of the laser beam is destroyed. This is important because even with the modern sophisticated image processing tools, analysis and interpretation of such images, in terms of morphology or velocity estimation for example, are very complicated due to the presence of these artifacts. As is shown in the following section of this article, the images obtained with the low coherence illumination are much more cleaner and sharper compared to the images obtained with direct laser illumination.

\begin{figure}[htb]
\centering
\includegraphics[width=0.45\columnwidth]{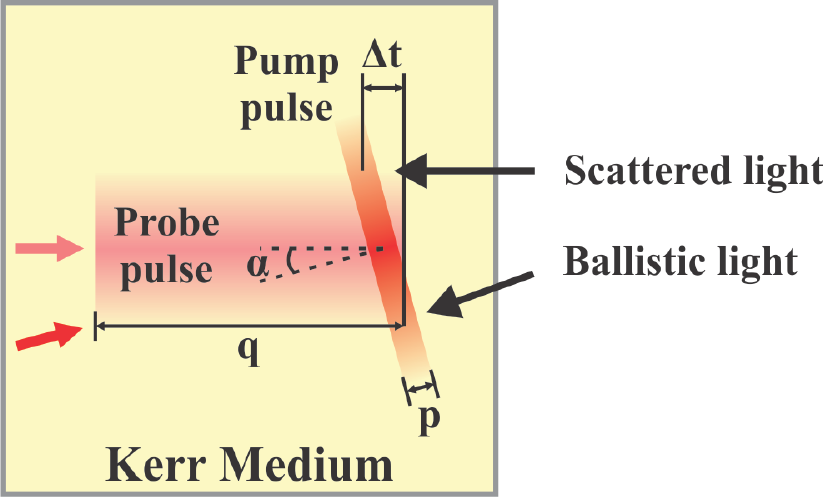}
\caption{Representation of one of the major problems with the crossed-beam geometry of the OKE-based time-gate due to non-collinear overlap of the pump and probe beams. $p$ and $q$ represents the pulse widths of the pump and probe pulses respectively and $\alpha$ is the angle between them.}
\label{fig:kerr}
\end{figure}

Another issue arises due to the non-collinear overlap of the pump (or switching) beam and the probe (or imaging) beam. In the classical crossed-beam OKE-based time-gates the angle between the pump and the probe beams not only influences optical gate's temporal characteristics but also restricts the use of OKE-based time-gates in time-resolved imaging \cite{Purwar2014,Idlahcen2009}. Due to the angle between the pump and the probe beams, the overlap between them does not occur at the same time. The diagram in Fig.~\ref{fig:kerr} represents the same and as a result the obtained time-resolved images do not correspond to the same time from the left to the right side of the image.

Here, we present an approach with the collinear incidence of the pump and probe beams at the Kerr medium for ultrafast time-gated imaging, using a low coherence, ultra-short source derived from supercontinuum (SC), generated using a femtosecond laser, for illumination. Introducing a low coherence illumination source limits the above mentioned artifacts keeping the pulse width short enough as demanded by the gating applications and the collinear overlap takes care of most of the issues with the OKE-based time gating.

\section{Supercontinuum generation}
\label{sec:SCgen}
\begin{figure}[htb]
\centering
\includegraphics[width=0.9\columnwidth]{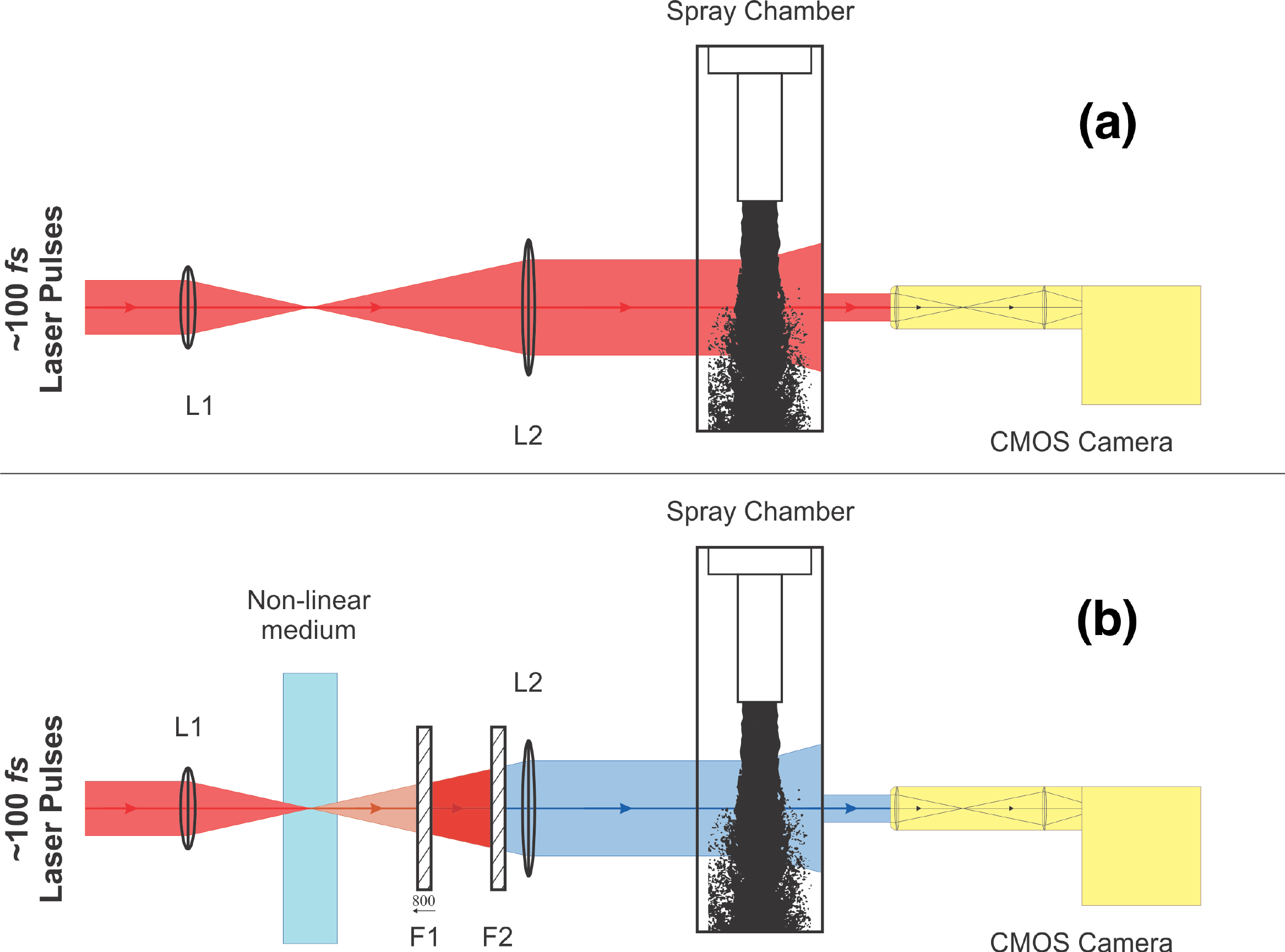}
\caption{Schematic of the experimental setup for in-line shadowgraphy with (a) direct laser and (b) SC-derived illuminations.}
\label{fig:setup2}
\end{figure}

A basic in-line shadowgraphy optical arrangement is shown in Fig.~\ref{fig:setup2} with direct laser and supercontinuum derived illuminations. The femtosecond laser pulses were generated using a regenerative amplifier (Libra, Coherent) seeded with a Titanium-Sapphire mode-locked laser (Vitesse, Coherent). The system is capable of generating laser pulses with pulse width less than $120$~fs and energy of about $3.5$~mJ per pulse at a repetition rate of $1$~kHz, pulse spectrum centered at wavelength $\lambda=800$~nm (fwhm $=12$~nm). The magnification of the optical setup can be adjusted with the help of lenses L1 (focal length, $f=100$~mm) and L2 ($f=200$~mm). 

On the other hand, supercontinuum was generated by tightly focusing these fs laser pulses inside a $4$~cm quartz cuvette filled with distilled water \cite{Durand2013,Nagura2002,Liu2002a,Brodeur1999}. A notch filter (F1) was used to filter out the fundamental laser ($\lambda=800$~nm) and a band-pass filter (F2) was used to select a small band of wavelengths for illumination of the fuel spray. A CMOS camera (LaVision Imager sCMOS) was used for recording the shadowgraphs.

\begin{figure}[htb]
\centering
\includegraphics[width=0.45\columnwidth]{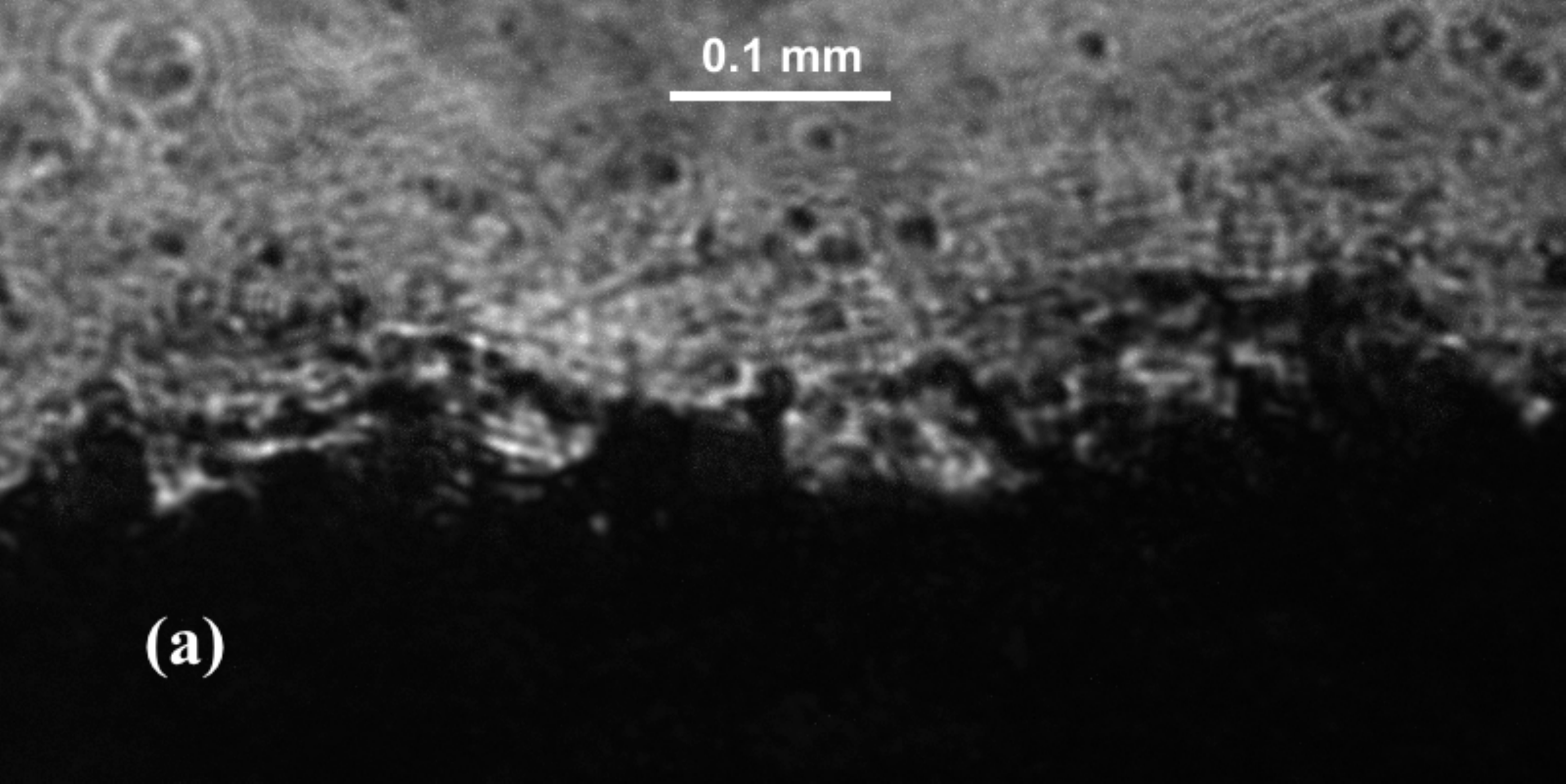} \quad
\includegraphics[width=0.45\columnwidth]{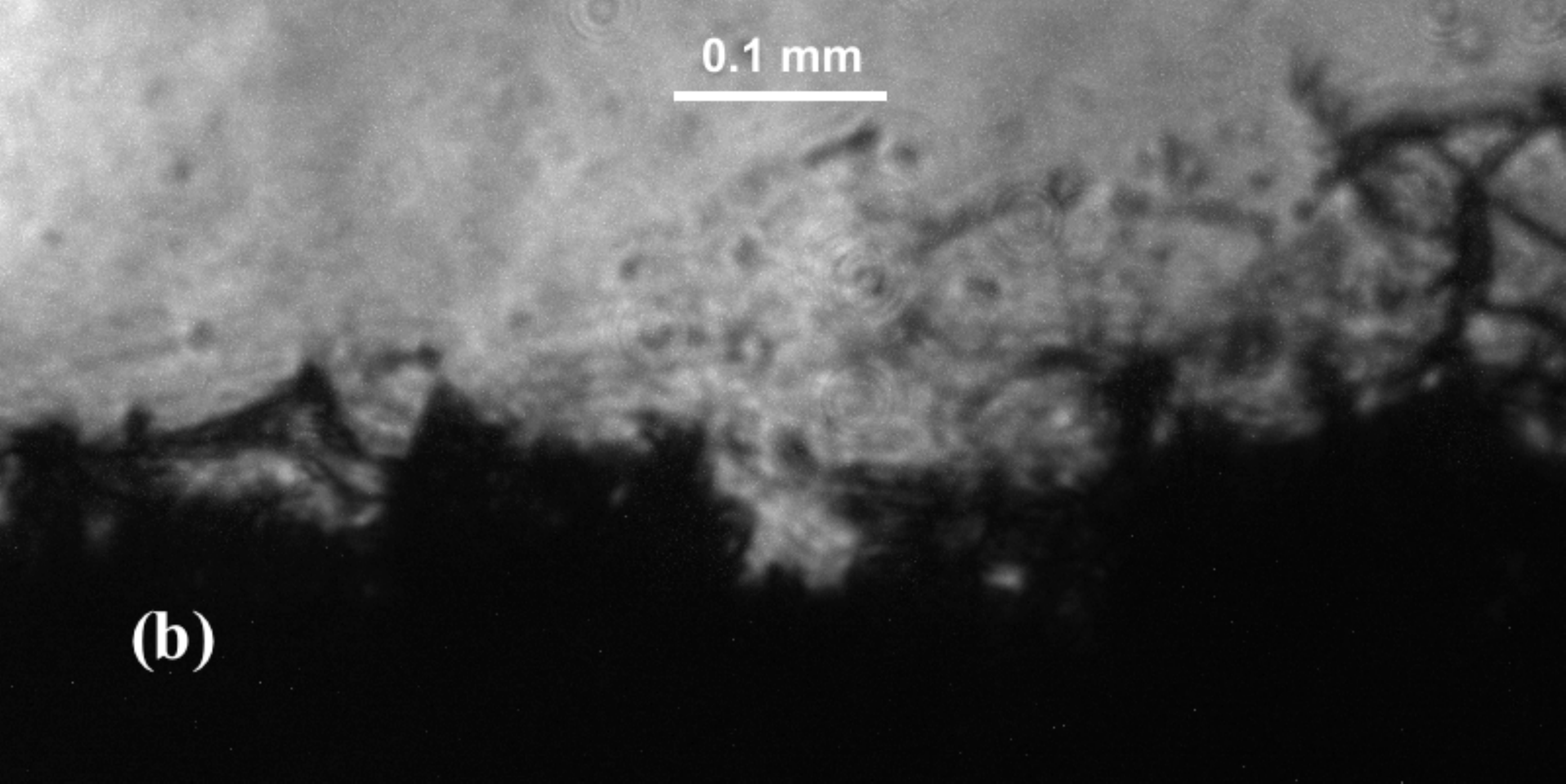}
\caption{A zoomed section of the spray images obtained using (a) direct laser and (b) supercontinuum derived illumination for qualitative comparison.}
\label{fig:zoom}
\end{figure}

This very simple experiment (Fig.~\ref{fig:setup2}) without any optical time gating showed, as was expected, that the quality of the spray images is highly improved by using SC-derived illumination. Figure~\ref{fig:zoom} shows a magnified view of a section from the diesel spray images obtained directly by illuminating the spray (a) with the coherent laser pulse and (b) with the light pulse extracted from SC, in the similar atmospheric conditions of pressure and temperature. The injection pressure for both these cases was constant at $400$~bars and the injector used was a single orifice, with nozzle diameter of about $180$~$\mu$m. Clearly, in the image obtained with SC-derived illumination, the unwanted artifacts of speckles from the laser are reduced to a great extent. Also note that the duration of the continuum pulse is still short enough to freeze the motion of the spray on the detector.

\subsection{Spectral and coherence measurements}
\begin{figure}[htb]
\centering
\includegraphics[width=0.9\columnwidth]{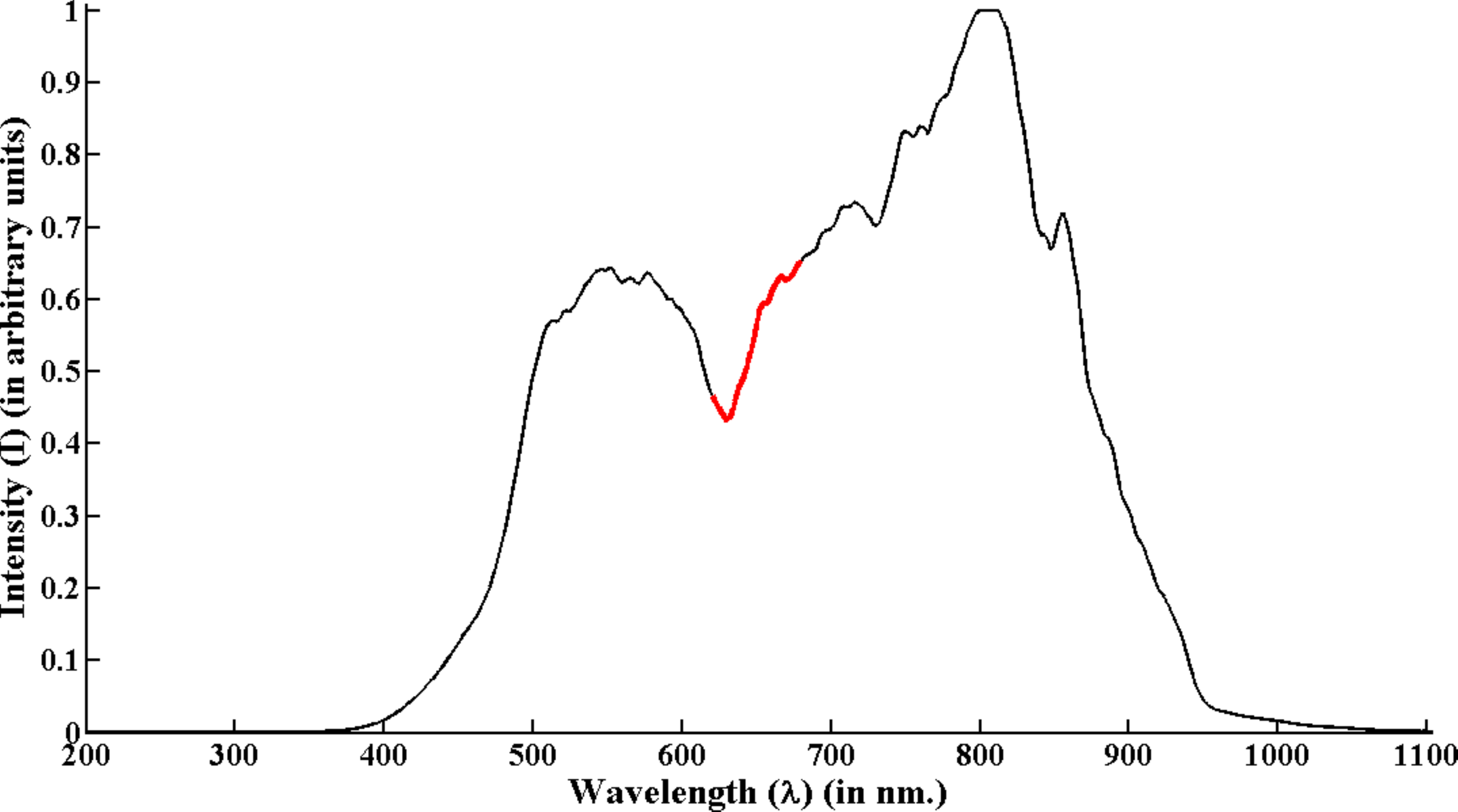}
\caption{Supercontinuum spectrum and the selected wavelength range used for illumination of the object under study with the optical gate configuration (curve highlighted in red, $\lambda=620-680$~nm).}
\label{fig:spectra}
\end{figure}

Before we use the SC derived light for time-resolved imaging it is important to characterize it in terms of spectral and coherence measurements. The spectrum of the SC, generated by focusing $800$~nm intense laser beam inside a water filled $4.0$~cm cuvette, was measured by a fiber-optic spectrometer (Ocean Optics Maya2000 Pro) and ranges from visible to near infrared region ($400-1100$~nm) as can be seen in Fig.~\ref{fig:spectra}. A small band of wavelengths ($\lambda=620-680$~nm) was selected from this continuum to be used as the probe beam in the optical time-gate setup for illumination of the object under study and is highlighted in red in Fig.~\ref{fig:spectra}.

\begin{figure}[htb]
\centering
\includegraphics[height=0.4\columnwidth]{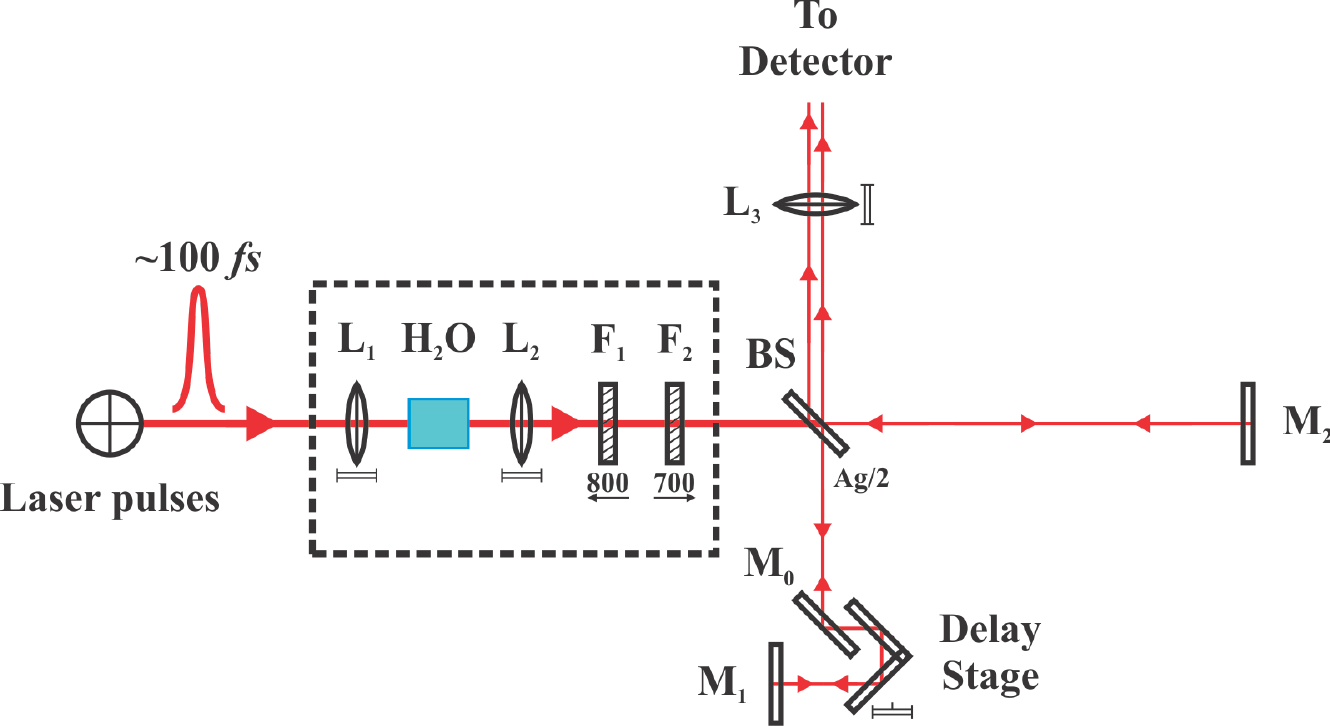}\quad 
\includegraphics[height=0.4\columnwidth]{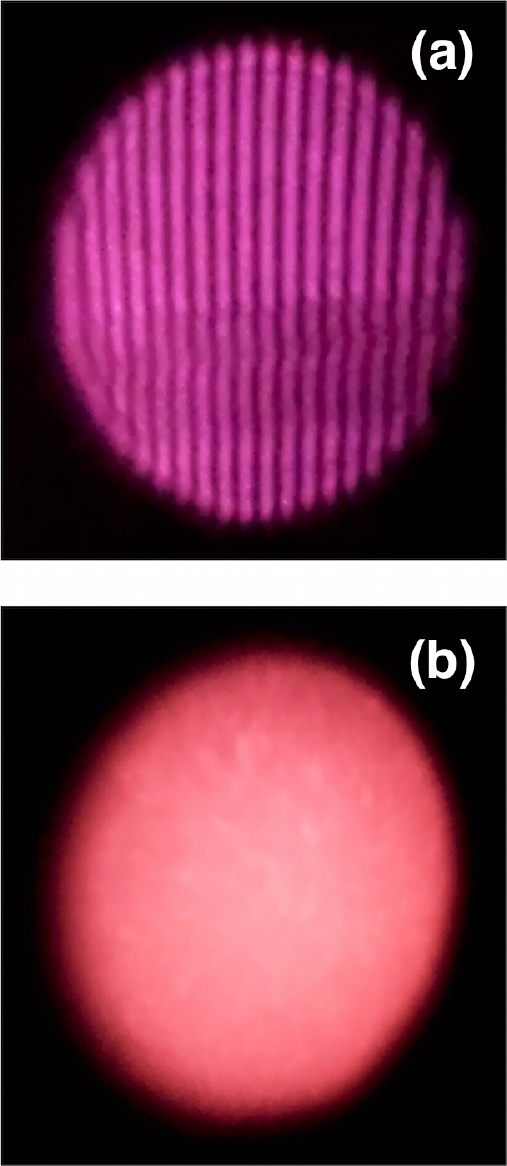}
\caption{Schematic of the Michelson's interferometer setup for measuring temporal coherence (left) and obtained interference patterns (right) for (a) fs laser and (b) SC-derived sources.}
\label{fig:setup_michel}
\end{figure}

The temporal coherence of the generated SC pulses were estimated using a Michelson's interferometer setup and were compared with the fs laser (original source used to generate the SC). A schematic of the experimental setup is shown in Fig.~\ref{fig:setup_michel}. A band of $13$~nm centered at $\lambda=700$~nm was chosen from the SC using the narrow band-pass filter $F_2$ for coherence measurement. Although, note that any chosen band of wavelengths with the same width, would have given a similar result.

Figure~\ref{fig:setup_michel} also shows the obtained interference patterns from the Michelson's interferometer setup with direct laser (top) and with the supercontinuum-derived source (bottom). It is very clear from these images that the temporal coherence of the laser beam has almost been destroyed at least to an extent that there is no detectable interference pattern. Also note that since there is almost no temporal coherence for a very narrow band of wavelengths, temporal coherence for a broader band would be even less since coherence length is inversely proportional to the bandwidth.

\begin{figure}[htb]
\centering
\includegraphics[width=0.8\columnwidth]{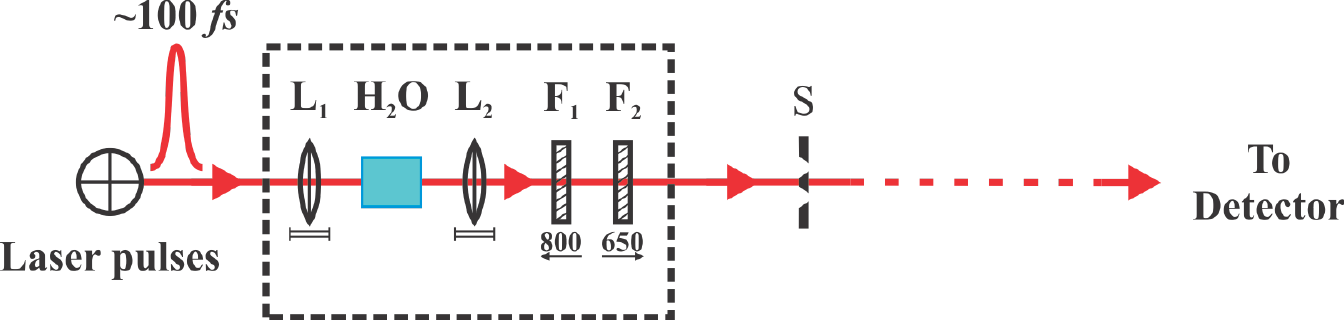}\\ \vspace{0.4cm}
\includegraphics[width=0.4\columnwidth]{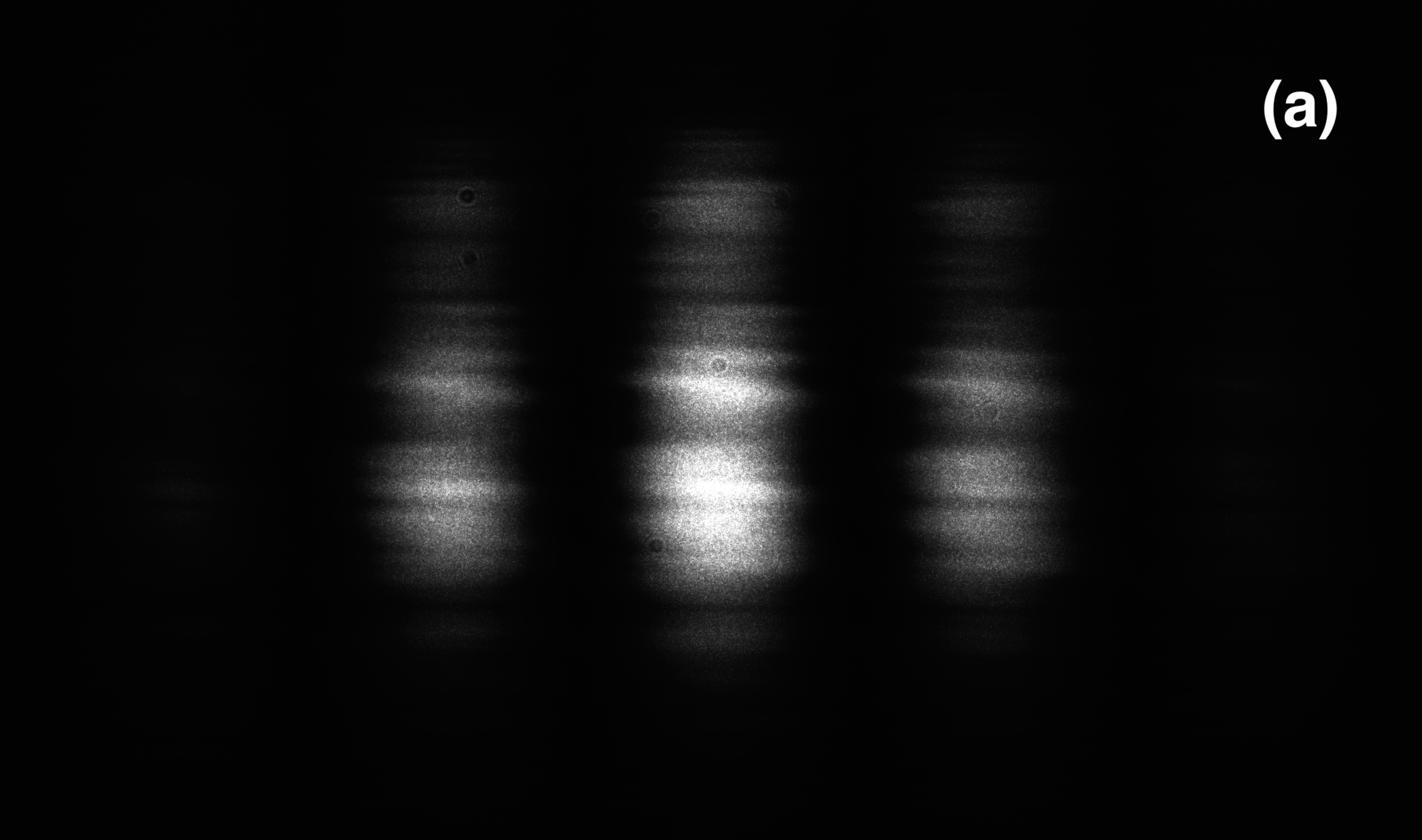}\quad \includegraphics[width=0.4\columnwidth]{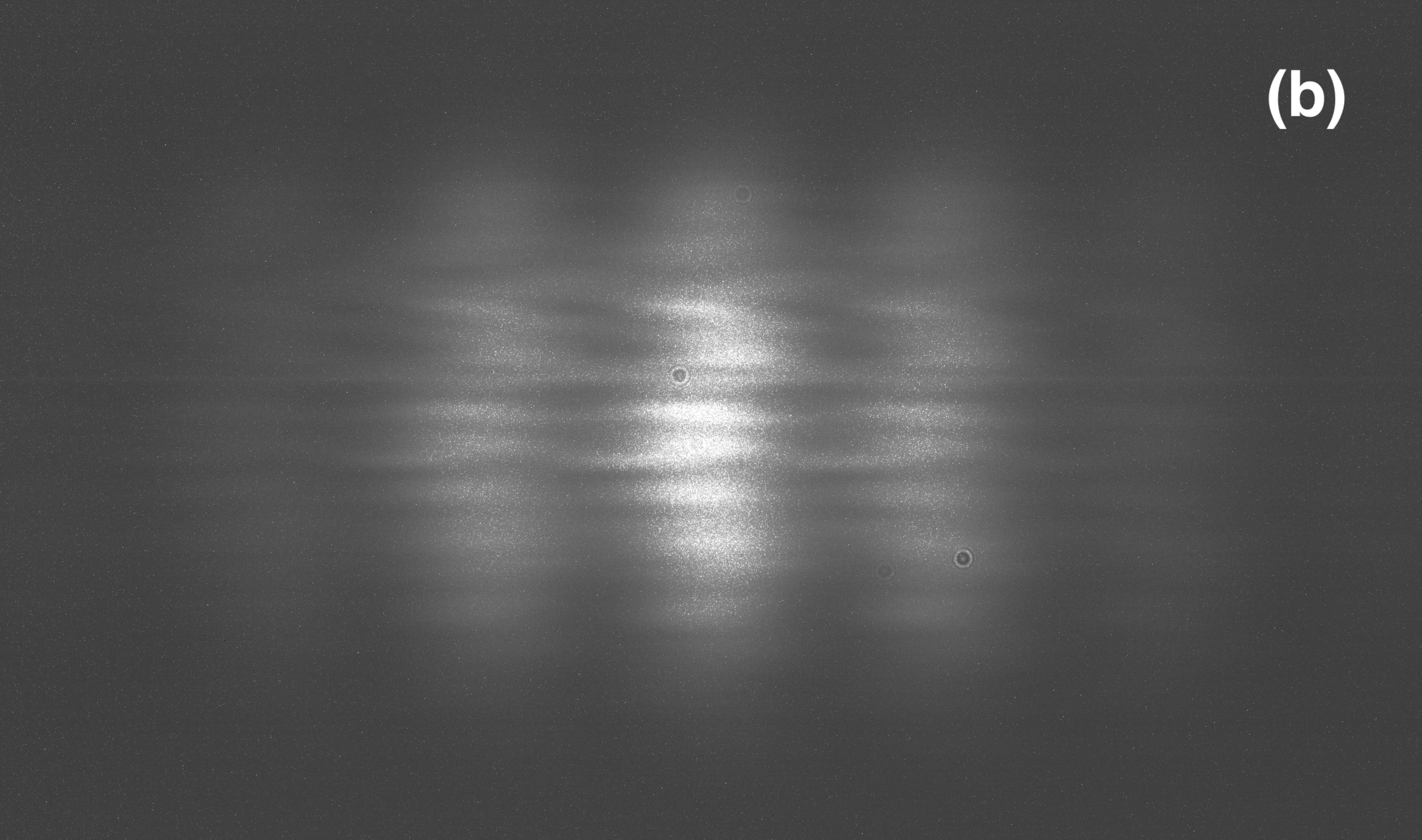}
\caption{Schematic of the Young's double-slit experimental setup for measuring spatial coherence (top) and obtained interference patterns (bottom) for (a) fs laser and (b) SC-derived illuminations.}
\label{fig:setup_young}
\end{figure}

On the other hand, spatial coherence of the SC-derived pulses was estimated using the Young's double slit experiment and the results were compared with the original source i.e. the fs laser. A schematic of the experimental setup is shown in Fig.~\ref{fig:setup_young} along with the obtained interference patterns. The width of the double slits $S$ used in these measurements was about $68$~$\mu$m and the distance between them was $428$~$\mu$m. Since spatial coherence is independent of the bandwidth of the light, apart from the fact that there might be an overlap of the interference patterns for various wavelengths, a larger band of $60$~nm centered at $\lambda=650$~nm was chosen to have enough intensity for detection.

From the above results it is safe to conclude that the SC-derived light is in fact low coherence light compared to the original fs laser source. While SC generation, which is a result of many coupled nonlinear processes such as filamentation, self-phase modulation, self-focusing, etc., \cite{Alfano2006} inside water, the temporal coherence of the beam is destroyed beyond detection whereas it still remains partially spatially coherent. Thus, unwanted laser speckles are significantly reduced in the spray images obtained with SC-derived illumination Fig.~\ref{fig:zoom}(b), due to destruction of the coherence of laser beam to a great extent.

\section{Collinear OKE-based time-gate, with SC-derived probe}
\label{sec:opticalgate}
\subsection{Optical setup}
\begin{figure}[htb]
\centering
\includegraphics[width=0.9\columnwidth]{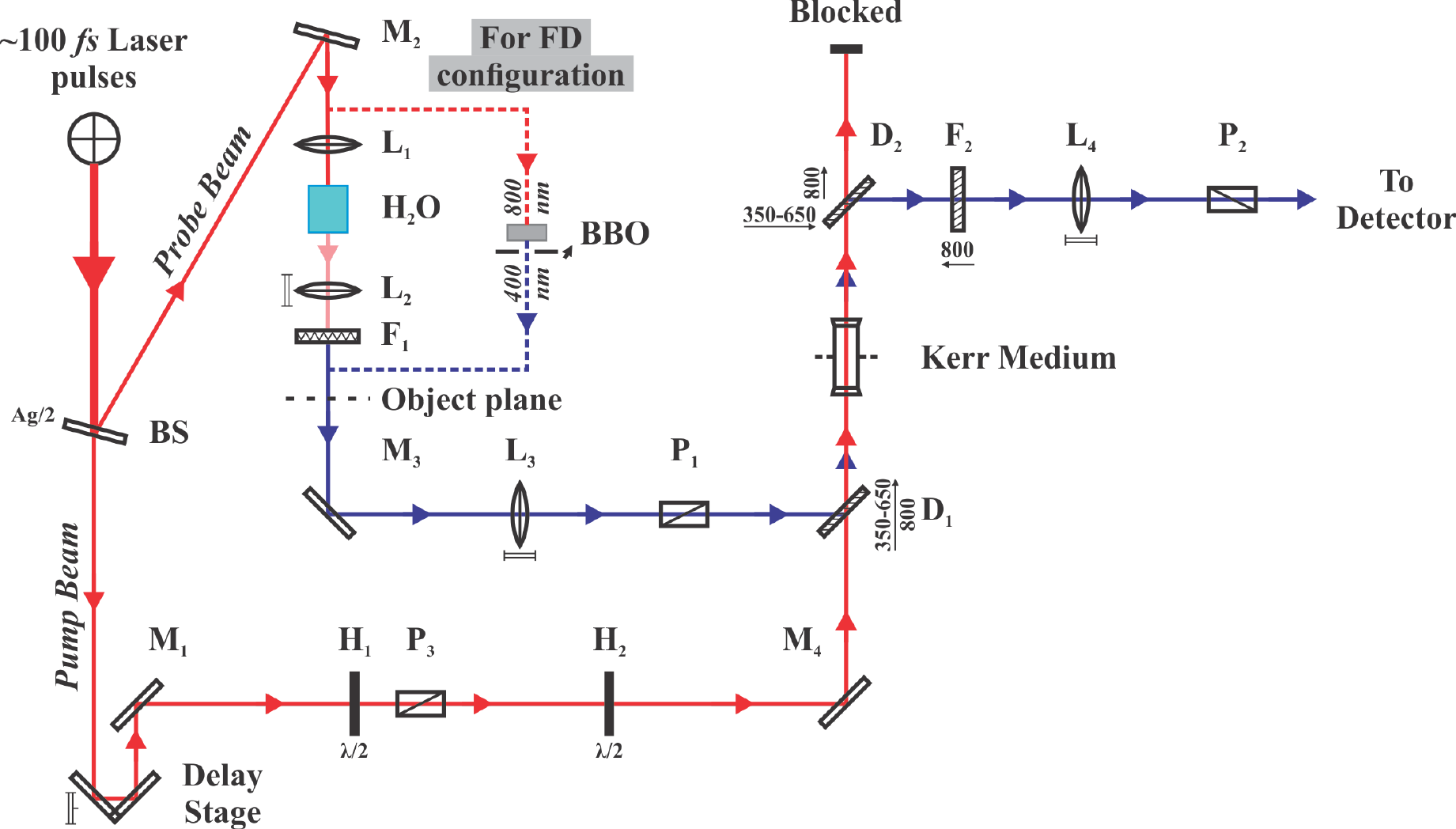}
\caption{Schematic of the experimental setup for ballistic imaging with low coherence, ultra-short supercontinuum-derived illumination and collinear incidence of the pump and probe beams inside the Kerr medium.}
\label{fig:setup}
\end{figure}

A schematic of the experimental setup for ballistic imaging of high-pressure fuel sprays with a low coherence, ultra-short pulsed illumination using an ultrafast optical Kerr effect based time-gating in collinear configuration is shown in Fig.~\ref{fig:setup}.

The incoming femtosecond laser beam was separated into probe and pump beams using a $50:50$ beam splitter (BS). The pump pulse ($\lambda=800$~nm), used to induce a transient birefringence in the Kerr medium (liquid carbon disulfide) for a duration depending on the non-linear properties of the medium \cite{Eichert2014} and on the pump pulse characteristics, passes through a computer controlled delay stage with a least count of $0.1$~$\mu$m (corresponding to a temporal delay of $0.67$~fs in air). The delay between the pump and the probe pulses is adjusted so that both pulses arrive at the Kerr medium at the same time and a good alignment takes care of their spatial overlap. In order to adjust the power of the pump beam a combination of a half waveplate (H$_1$) and a polarizer (P$_3$) is used. Another half waveplate (H$_2$) is used to rotate the polarization axis of the pump by an angle, with respect to the probe beam's polarization, which maximizes the induced birefringence in the Kerr medium, thus maximizing the efficiency of the time-gate.

The probe pulse is extracted from the supercontinuum generated by tightly focusing the high power femtosecond laser pulses inside a 4 cm cuvette filled with distilled water using a short focal length biconvex lens L$_1$. An appropriate band pass filter (F$_1$, central $\lambda=650$~nm fwhm $=60$~nm) extracts a narrow range of wavelengths for illumination of the fuel spray and at the same time also suppresses the incoming fundamental wavelength ($\lambda=800$~nm). An appropriate notch filter could also be used to block the incoming fundamental wavelength if necessary. It is now known that the pulse width for individual wavelengths after supercontinuum generation does not change significantly \cite{Alfano2006} even if the overall pulse width may increase due to group velocity dispersion, as light propagates through its course. The narrow band pass filter though reduces this artifact. The probe pulse then illuminates the object under study.

Since the pump and the probe beams here have different wavelengths, it is possible to combine them before the Kerr medium and to separate them afterwards using the dichroic filters (D$_1$ and D$_2$). Using this principle both pulses are incident at the Kerr medium in a collinear fashion. The dichroic filters used here were Brightline single-edge dichroic filters with average reflection greater than $95$\% for $\lambda=350-650$~nm and average transmission greater than $93$\% for $\lambda=673.7-950$~nm. After interaction with the object, the probe pulse passes through the Kerr medium, which is sandwiched between two crossed polarizers P$_1$ and P$_2$. A notch filter (F$_2$, $\lambda=808$~nm, fwhm $=41$~nm) has been used to filter out any remaining fundamental wavelength ($\lambda=800$~nm) from probe beam (due to the limitations of the dichroic filters) or pump beam (due to its scattering through the Kerr medium) from reaching the detector. An additional neutral density filter may be used to adjust the intensity of light reaching the detector according to its limitations. For imaging purposes a scientific CMOS camera (LaVision Imager sCMOS) was used as a detector whereas for basic intensity measurements a power meter (Ophir Nova-II) was used.

\subsection{Spatial and temporal resolution measurements}
\begin{figure}[htb]
\centering
\includegraphics[width=0.9\columnwidth]{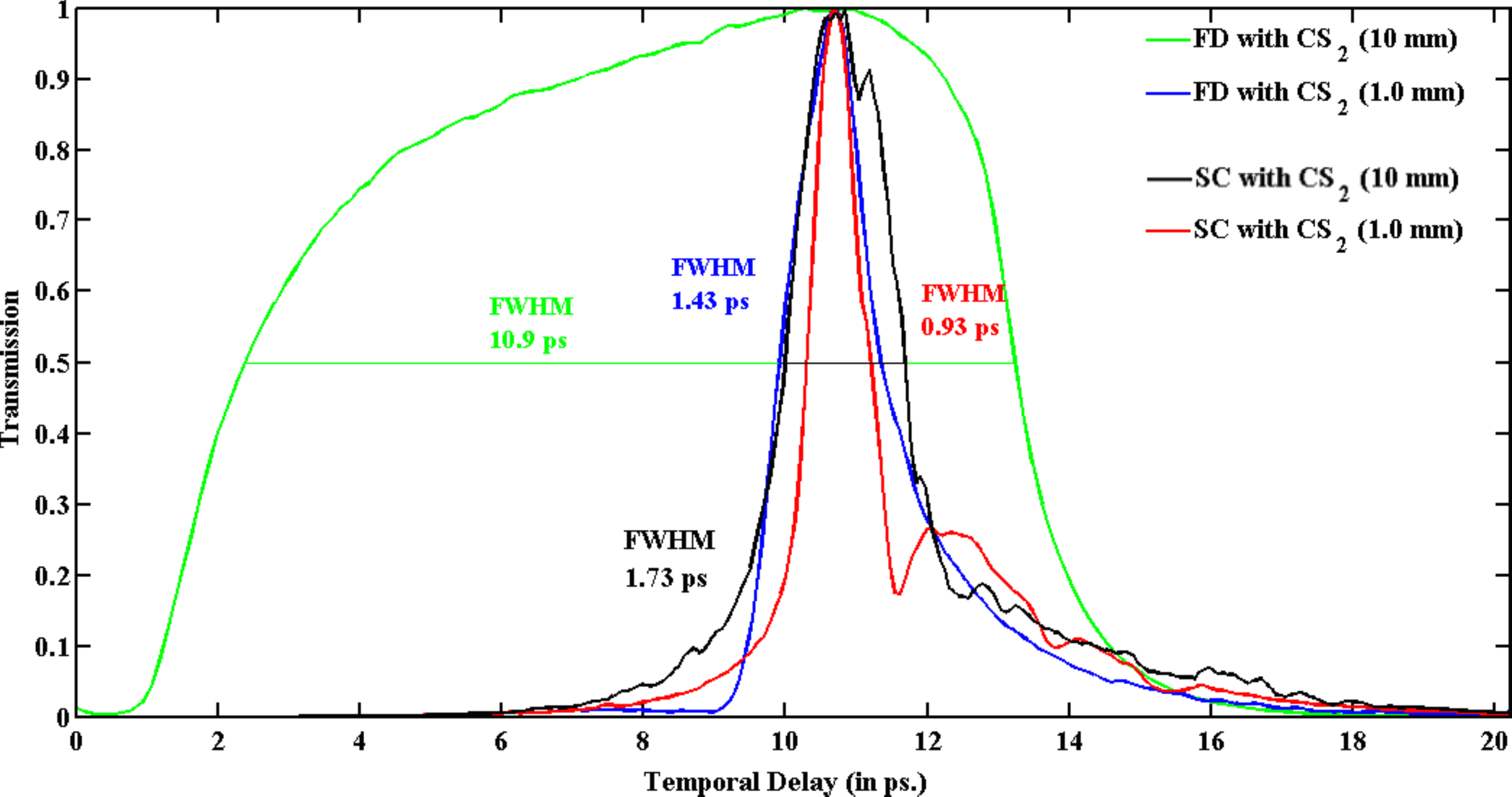}
\caption{The temporal profile of the OKE-based time gate (with $1.0$~mm and $10$~mm CS$_2$ as Kerr medium) for SC-derived probe pulse ($\lambda=635$~nm, in red and black) and for frequency-doubled (FD) probe ($\lambda=400$~nm, in blue and green).}
\label{fig:temporal}
\end{figure}

The temporal resolution and time profile of the Kerr-effect based optical gate was extracted from the averaged spectra measured using the aforementioned spectrometer by varying the delay between the pump and the probe pulses with a delay stage as shown in Fig.~\ref{fig:setup}. The temporal profiles for the optical gate with $1.0$~mm and $10$~mm thick Kerr medium i.e. liquid carbon disulfide (CS$_2$) with SC-derived probe beam $\lambda=620-650$~nm (band reduced further due to rejection by the dichroic filters) along with the frequency doubled probe (using the principle of second harmonic generation, $\lambda=400$~nm) is shown in Fig.~\ref{fig:temporal} for comparison.

\begin{figure}[htb]
\centering
\includegraphics[width=0.9\columnwidth]{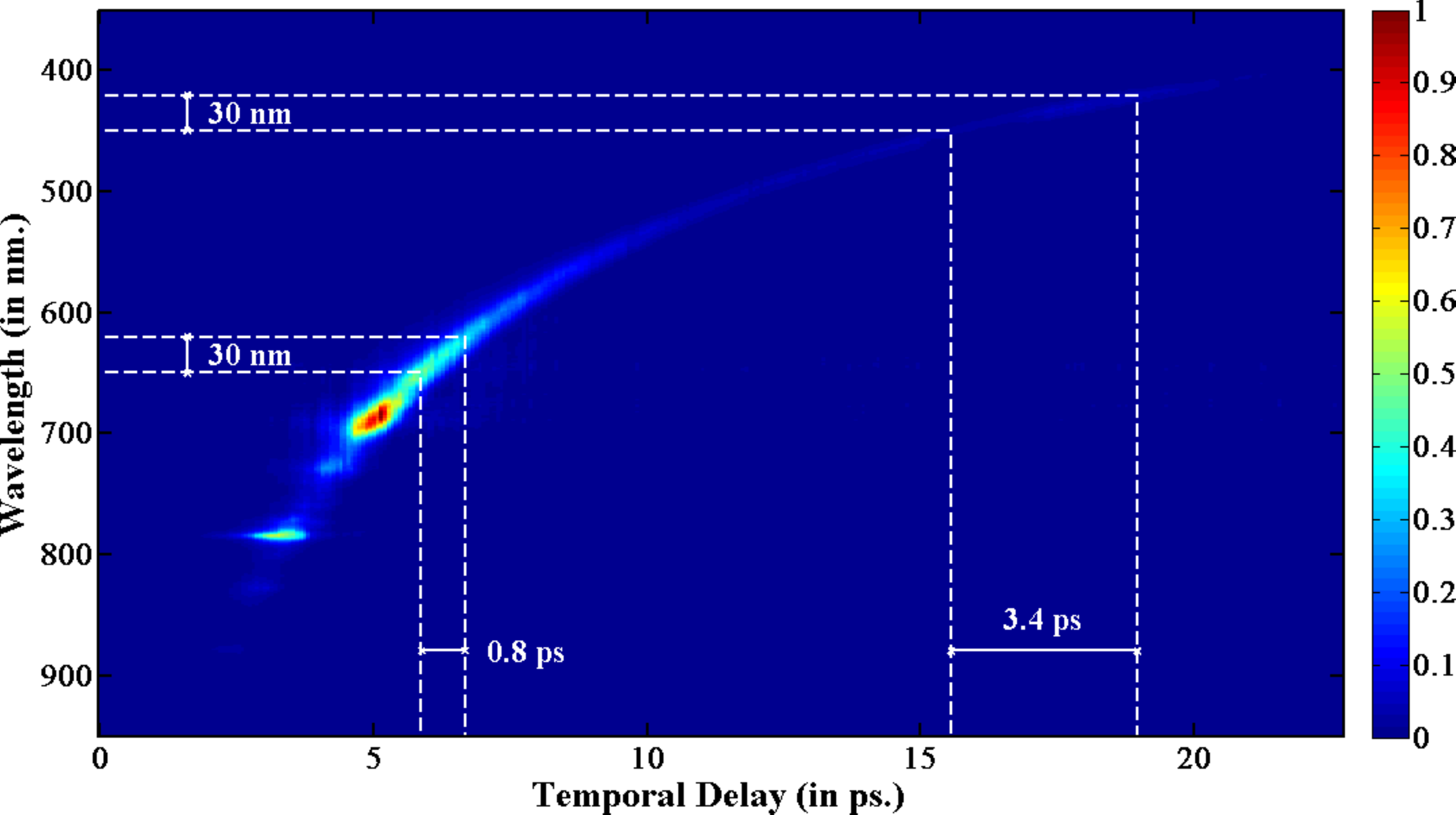}
\caption{The chirp characteristics for the entire SC. The abscissa shows the temporal delay between pump and probe pulses at the Kerr medium, $1.0$~mm liquid CS$_2$.}
\label{fig:chirp}
\end{figure}

The optical gate durations with SC-derived probe ($\lambda=635$~nm) for $1.0$~mm and $10$~mm thick CS$_2$ cells are $0.93$~ps and $1.73$~ps respectively, whereas the same for frequency-doubled probe ($\lambda=400$~nm) are $1.43$~ps for $1.0$~mm thick CS$_2$ cell and $10.9$~ps for $10$~mm. The huge variation in the optical gate duration for $10$~mm thick Kerr medium is due to the group velocity dispersion inside liquid CS$_2$. The group velocity for $\lambda=400$~nm ($v_g^{400}$) in liquid CS$_2$ is lower than the group velocity for $\lambda=800$~nm ($v_g^{400}<v_g^{800}$). Hence, the $800$~nm pulse (pump) can catch up and overlap with the $400$~nm probe pulse while traversing through the liquid CS$_2$. Note that this overlapping has no effect if it occurs outside the Kerr medium, but acts on the polarization of the probe pulse when it occurs inside the Kerr medium. Limiting the thickness of the CS$_2$ cell then reduces the possibility of overlapping of the two pulses inside the cell and diminishes the optical gate duration. The same happens for the $\lambda=635$~nm SC-derived pulse and $\lambda=800$~nm pump pulse, but for these wavelengths the difference between the group velocities in CS$_2$ is small and the gate duration does not increase so drastically when the thickness of the CS$_2$ cell is increased from $1.0$~mm to $10$~mm. This allows us to use a broader band of wavelengths near $800$~nm so as to have sufficient light for illumination and detection without affecting the optical gate duration significantly. It is easy to comprehend this from Fig.~\ref{fig:chirp} showing the dependence of the gate duration on the chosen wavelength-band from the generated SC taking into account the effect of group velocity dispersion in liquid CS$_2$ \cite{Nagura2002}. Note that the chirp characteristics for the entire SC were measured with the crossed-beam (or non-collinear) configuration of the OKE-based time-gate due to the limitations of the dichroic mirrors in the collinear configuration: the non-filtered SC spreads inside the CS$_2$ cell depending on the group velocities for individual wavelengths and then the delay between pump and SC pulses selects a narrow band of wavelengths, which is measured by the spectrometer, giving Fig.~\ref{fig:chirp}.

Spatial resolution, on the other hand is estimated using the normalized modulation transfer function (MTF), which, for the presented optical setup (Fig.~\ref{fig:setup}) was measured using the slanted-edge method (ISO 12233 standard for MTF measurement of electronic still-picture cameras \cite{patent}). The MTF can be calculated directly from the edge-spread function (ESF) using,
\[\text{MTF}(\omega)=\int{\frac{d\text{ESF}(x)}{dx} e^{-2i\pi\omega x}dx}\]
where $\omega$ is the spatial frequency in lines/mm, $x$ is the abscissa along an axis orthogonal to the slanted edge. In other words, MTF is the Fourier transform of the derivative of the ESF, which is also the line-spread function. It should be noted that while calculation of MTF from the over-sampled ESF data, the differentiation along the slanted edge is very sensitive to the high-frequency noise and in order to avoid amplifying this noise during the computation of the derivative, the averaged over-sampled ESF is fitted using the Logistic (Fermi) function $f(x)$ \cite{Li2009},
\[\text{ESF}(x)=d+f(x)=d+\frac{a}{1+\exp{[(x-b)/c]}}\]
where $a$, $b$, $c$ and $d$ are the desired fit parameters.

\begin{figure}[htb]
\centering
\includegraphics[width=0.9\columnwidth]{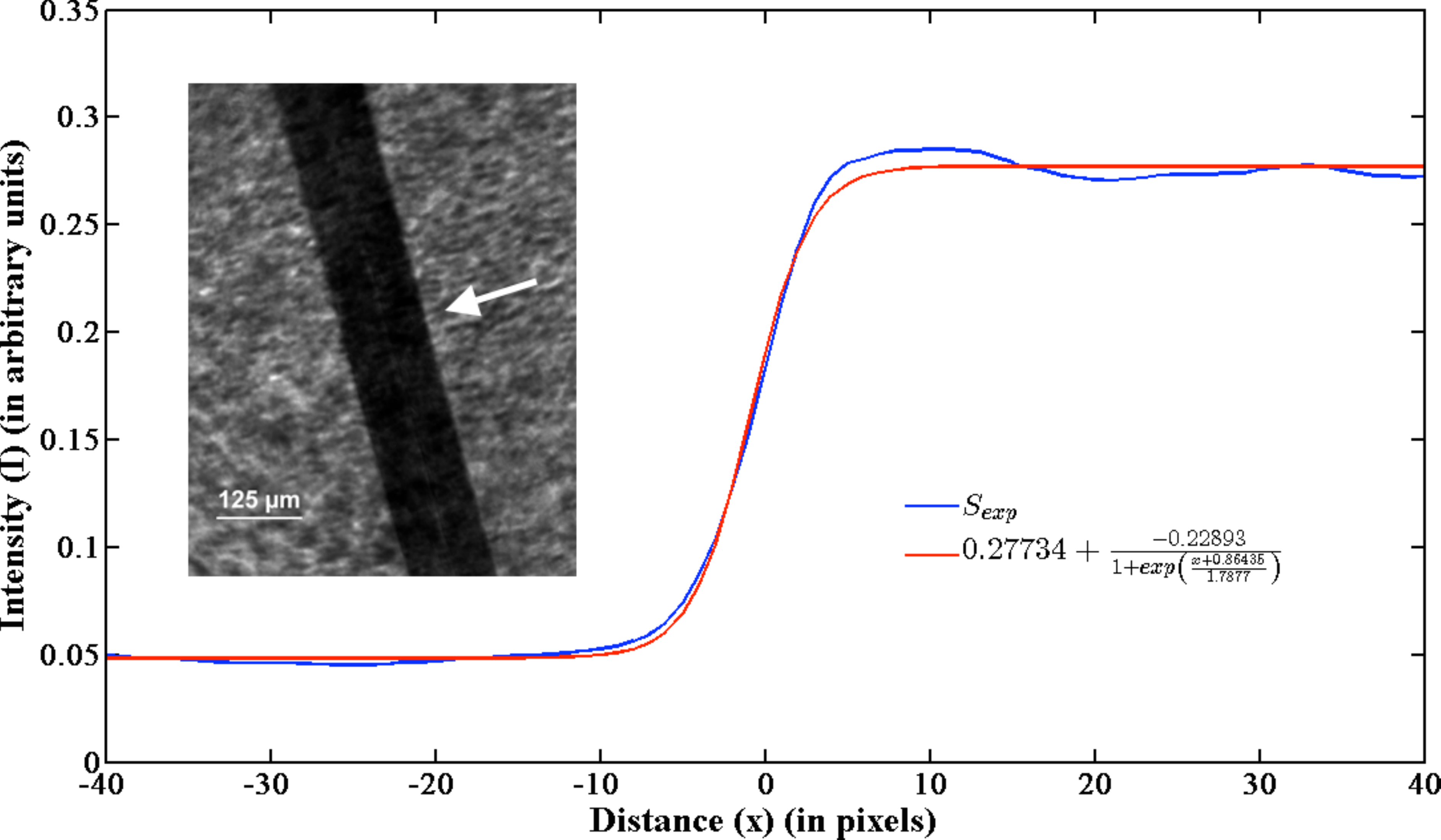}
\caption{Image of a slanted glass fiber ($\phi=125$~$\mu$m) obtained using SC-derived illumination with collinear optical time-gate configuration and the corresponding averaged ESF extracted from the indicated edge in the image (blue, solid line curve) and fitted ESF (red, solid line curve).}
\label{fig:esf}
\end{figure}

\begin{figure}[htb]
\centering
\includegraphics[width=0.9\columnwidth]{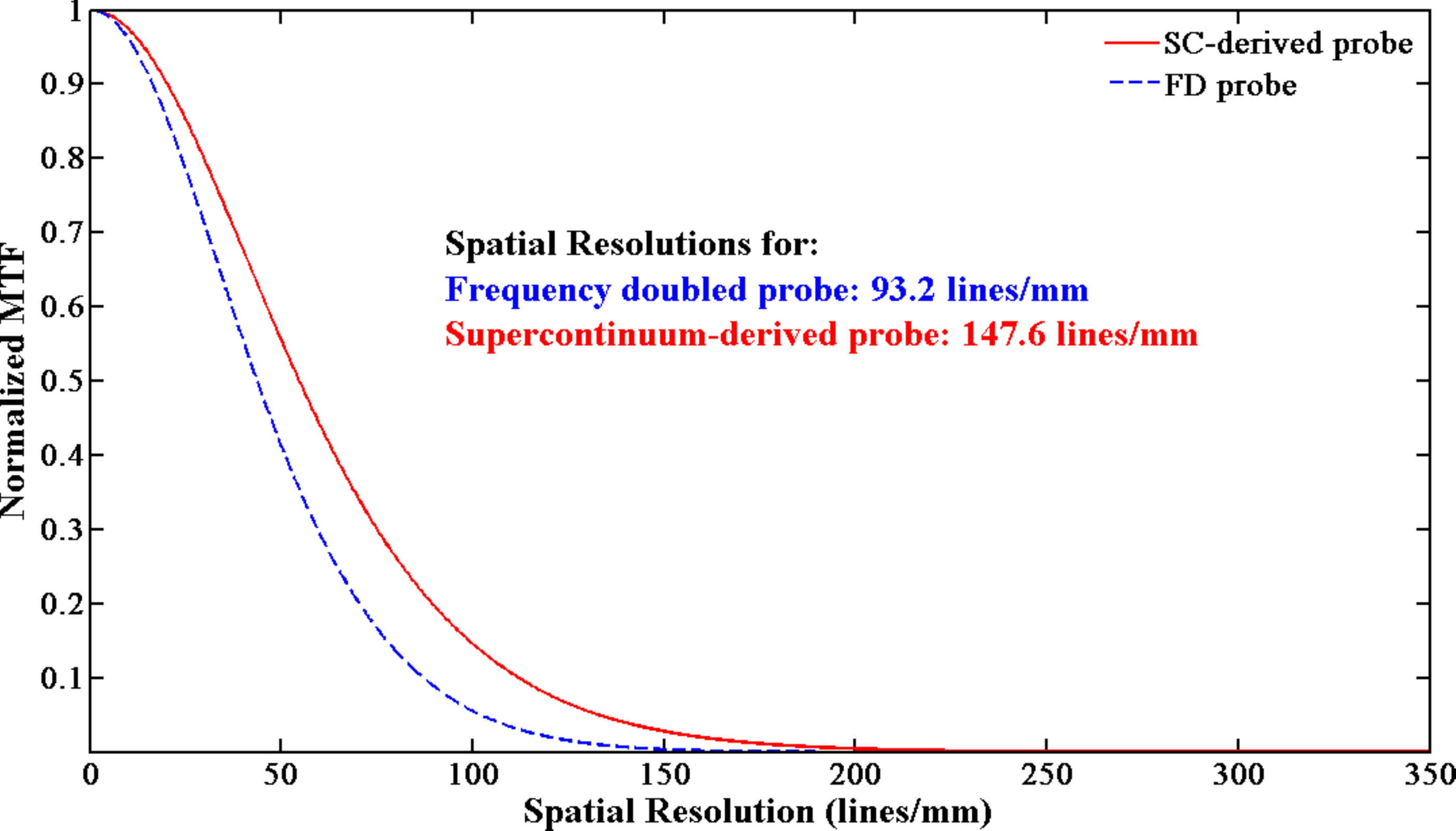}
\caption{The normalized modulation transfer function (MTF) as a function of spatial frequency ($\omega$) (in lines/mm) calculated using the fitted ESFs for the collinear configuration of the optical gate with SC-derived (red, solid line curve) and FD (blue, dotted line curve) probe beams for illumination.}
\label{fig:mtf}
\end{figure}

Figure~\ref{fig:esf} and \ref{fig:mtf} show the fitted ESF and normalized MTF calculated for the SC-derived probe, overlapping with the pump pulse in a collinear arrangement within the Kerr medium, placed at the image plane of the lens L$_3$. For comparison Fig.~\ref{fig:mtf} also shows the normalized MTF for frequency-doubled probe ($\lambda=400$~nm) obtained by second harmonic generation in a beta-barium borate (BBO) crystal \cite{Purwar2014}. The achievable imaging spatial resolution from the MTF curve (inverse of Nyquist frequency) with 1.0 mm liquid CS$_2$ as the Kerr medium for the supercontinuum-derived probe was found to be 147.6 lines/mm while for the frequency-doubled probe, in the same configuration, it was found to be 93.2 lines/mm.

\subsection{Application to high-pressure fuel sprays}
\begin{figure}[htb]
\centering
\includegraphics[width=0.47\columnwidth]{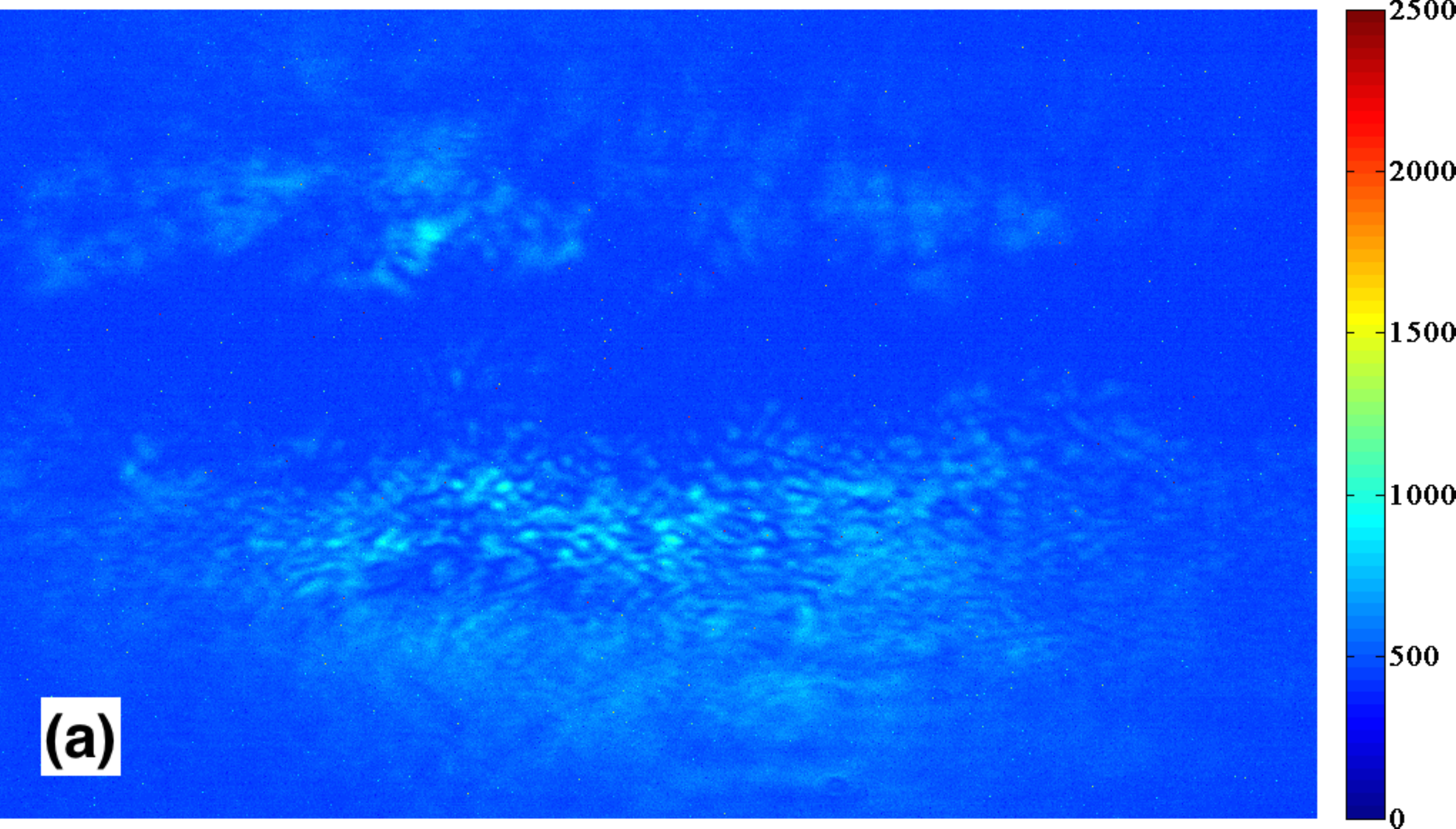}\quad
\includegraphics[width=0.47\columnwidth]{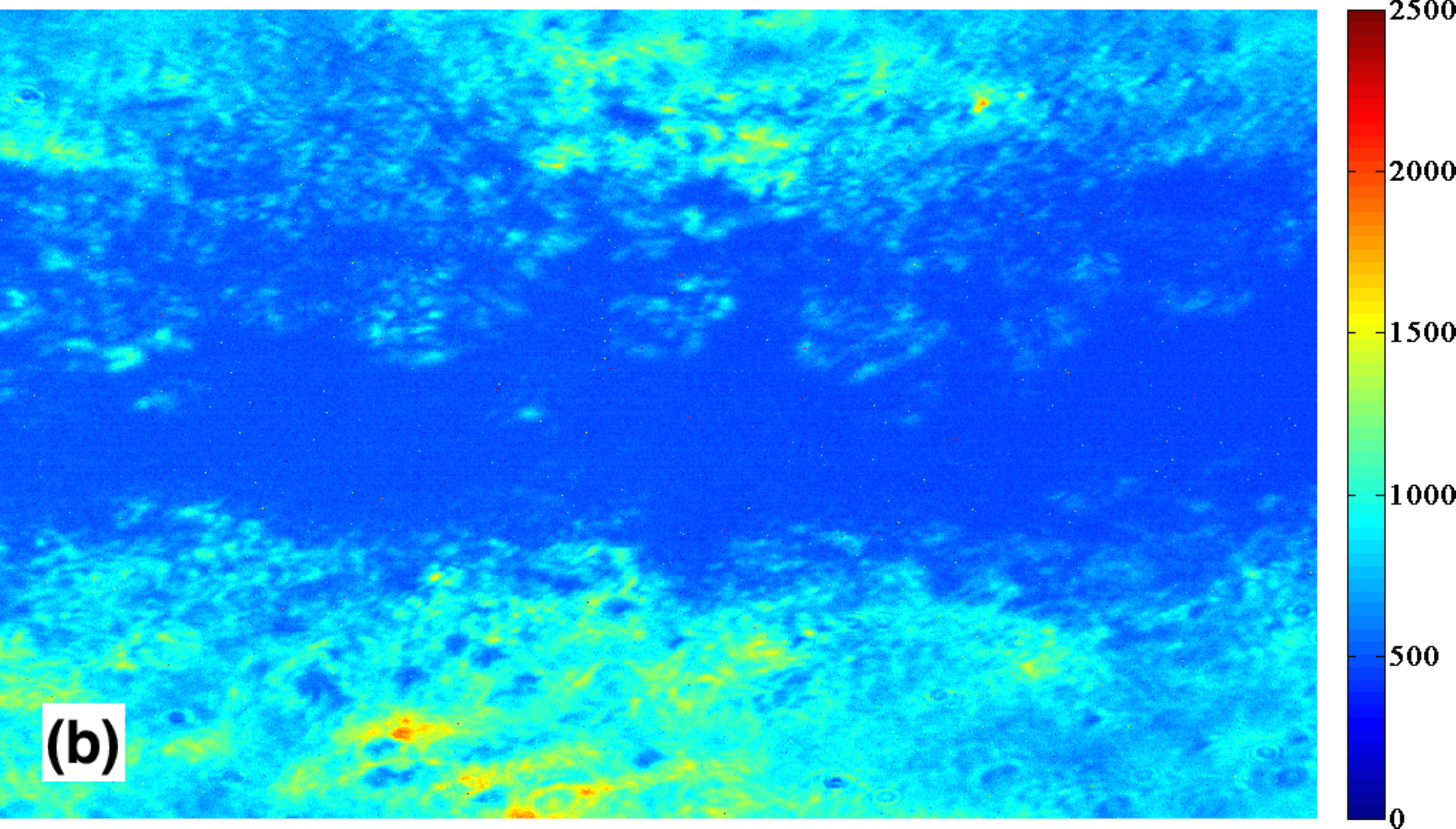}
\caption{Ballistic images of fuel spray with (a) frequency-doubled, coherent and (b) SC-derived, low coherence illumination sources with the collinear OKE-based time-gate.}
\label{fig:sprays}
\end{figure}

Finally after the successful development and thorough characterization of the optical gate using low coherence probe pulse derived from SC, it has initially been tested for the imaging of high-pressure fuel sprays. Figure~\ref{fig:sprays} shows a comparison of the spray images 4 cm away from the nozzle (single-orifice injector, nozzle diameter = 185 µm) in the direction of the flow, obtained with the collinear configuration of the optical gate with coherent (frequency doubled) and low coherence (SC-derived) illumination sources. The injection pressure was kept constant at around 400 bars for both these cases. The speckles due to the laser are clearly visible in the time-resolved images obtained by using coherent illumination whereas the same are largely reduced when the spray was illuminated by the supercontinuum derived, low coherence light.

\section{Conclusions}
\label{conclu}
An OKE-based time-gate is presented with the collinear incidence of the pump and the probe beams at the Kerr medium, 1.0 mm liquid CS$_2$, thereby solving several issues linked with the non-collinear overlap of the two in the classical configuration of the optical gate discussed in detail in Ref.~\cite{Purwar2014}. In order to remove the artifacts arising due to the coherence of the laser beam in the spray images, for example laser speckles, the coherence of the laser beam was partially destroyed by tightly focusing these femtosecond laser pulses inside a water cuvette, generating a wide band of white light continuum. Several non-linear optical phenomena are held responsible for this like self-phase modulation, self-focusing, etc. The coherence properties of the SC-derived source have been studied using the Michelson's interferometer setup (for temporal coherence) and Young's double-slit experiment (for spatial coherence). A very good interference pattern was obtained with the Michelson's interferometer using directly the fs laser but no detectable interference could be obtained with the SC-derived source suggesting that the SC-derived source is temporally incoherent. However, during SC generation the spatial coherence of the laser beam was not destroyed completely but was significantly reduced.

The proposed optical gate setup with low coherence, SC-derived probe was first characterized in terms of spatial and temporal resolution. The obtained results were compared with the previously proposed dual color, collinear optical-gate setup with coherent, frequency-doubled probe and it was found that the spatial resolution of the optical gate is significantly improved. An additional advantage of using SC-derived light as probe or imaging beam is that one can choose a thicker Kerr medium (CS$_2$ cell) for better efficiency of the time-gate and a narrow band of wavelengths as close as possible to the pump's wavelength i.e. $800$~nm from the SC, without significantly affecting the temporal resolution of the optical gate. Also since the pump and probe beams can be separated easily using notch and band-pass filters the noise due to the high-power pump beam is highly reduced in this collinear configuration of the OKE-based time-gate.

Furthermore a narrow band of wavelengths from SC were selected for illumination of the fuel spray. The time-resolved spray images obtained with the low-coherence illumination were found to be significantly better than those obtained with the coherent illumination using the collinear OKE-based time-gate of duration of about $1$~ps for both the cases.

\section*{Acknowledgments}
This work was supported by NADIA-Bio program, with funding from the French Government and the Haute-Normandie region, in the framework of the Moveo cluster (``private cars and public transport for man and his environment").

\begin{thebibliography}{10}
\newcommand{\enquote}[1]{``#1''}

\bibitem{Alfano1997}
R.~R. Alfano, S.~G. Demos, and S.~K. Gayen, \enquote{{Advances in Optical
  Imaging of Biomedical Media},} Ann. NY. Acad. Sci. \textbf{820}, 248--271
  (1997).

\bibitem{Wang1991a}
L.~Wang, Y.~Liu, P.~P. Ho, and R.~R. Alfano, \enquote{{Ballistic imaging of
  biomedical samples using picosecond optical Kerr gate},} Proc. SPIE
  \textbf{1431}, 97--101 (1991).

\bibitem{Linne2013a}
M.~Linne, \enquote{{Imaging in the optically dense regions of a spray: A review
  of developing techniques},} Prog. Energy Combust. Sci. \textbf{39}, 403--440
  (2013).

\bibitem{Berrocal2007}
E.~Berrocal, D.~L. Sedarsky, M.~E. Paciaroni, I.~V. Meglinski, and M.~a. Linne,
  \enquote{{Laser light scattering in turbid media Part I: Experimental and
  simulated results for the spatial intensity distribution.}} Opt. Express
  \textbf{15}, 10649 (2007).

\bibitem{Berrocal2009}
E.~Berrocal, D.~L. Sedarsky, M.~E. Paciaroni, I.~V. Meglinski, and M.~a. Linne,
  \enquote{{Laser light scattering in turbid media Part II: Spatial and
  temporal analysis of individual scattering orders via Monte Carlo
  simulation.}} Opt. Express \textbf{17}, 13792 (2009).

\bibitem{Duran2015}
S.~P. Duran, J.~M. Porter, and T.~E. Parker, \enquote{{Ballistic Imaging of
  Diesel Sprays Using a Picosecond Laser: Characterization and Demonstration},}
  Appl. Opt. (to be published).

\bibitem{Rahm2014}
M.~Rahm, M.~Paciaroni, Z.~Wang, M.~A. Linne, and D.~Sedarsky, \enquote{{Optical
  Arrangements for Time-Gated Ballistic Imaging},} Imaging Appl. Opt. 2014 p.
  IM4C.5 (2014).

\bibitem{Idlahcen2011}
S.~Idlahcen, C.~Roz\'{e}, L.~M\'{e}\`{e}s, T.~Girasole, and J.-B. Blaisot,
  \enquote{{Sub-picosecond ballistic imaging of a liquid jet},} Exp. Fluids
  \textbf{52}, 289--298 (2011).

\bibitem{Linne2005}
M.~a. Linne, M.~Paciaroni, J.~R. Gord, and T.~R. Meyer, \enquote{{Ballistic
  imaging of the liquid core for a steady jet in crossflow.}} Appl. Opt.
  \textbf{44}, 6627--34 (2005).

\bibitem{Wang1991}
L.~Wang, P.~Ho, C.~Liu, G.~Zhang, and R.~Alfano, \enquote{{Ballistic 2-d
  imaging through scattering walls using an ultrafast optical kerr gate.}}
  Science (80-. ). \textbf{253}, 769--771 (1991).

\bibitem{Purwar2014}
H.~Purwar, S.~Idlahcen, C.~Roz\'{e}, D.~Sedarsky, and J.-B. Blaisot,
  \enquote{{Collinear, two-color optical Kerr effect shutter for ultrafast
  time-resolved imaging},} Opt. Express \textbf{22}, 15778--15790 (2014).

\bibitem{Idlahcen2009}
S.~Idlahcen, L.~M\'{e}\`{e}s, C.~Roz\'{e}, T.~Girasole, and J.-B. Blaisot,
  \enquote{{Time gate, optical layout, and wavelength effects on ballistic
  imaging.}} J. Opt. Soc. Am. A \textbf{26}, 1995--2004 (2009).

\bibitem{Durand2013}
M.~Durand, K.~Lim, V.~Jukna, E.~Mckee, M.~Baudelet, A.~Houard, M.~Richardson,
  A.~Mysyrowicz, and A.~Couairon, \enquote{{Influence of the anomalous
  dispersion on the supercontinuum generation by femtosecond laser
  filamentation},} in \enquote{CLEO,}  (2013), pp. 1--2.

\bibitem{Nagura2002}
C.~Nagura, A.~Suda, H.~Kawano, M.~Obara, and K.~Midorikawa,
  \enquote{{Generation and characterization of ultrafast white-light continuum
  in condensed media.}} Appl. Opt. \textbf{41}, 3735--42 (2002).

\bibitem{Liu2002a}
W.~Liu, O.~Kosareva, I.~Golubtsov, a.~Iwasaki, a.~Becker, V.~Kandidov, and
  S.~Chin, \enquote{{Random deflection of the white light beam during
  self-focusing and filamentation of a femtosecond laser pulse in water},}
  Appl. Phys. B Lasers Opt. \textbf{75}, 595--599 (2002).

\bibitem{Brodeur1999}
A.~Brodeur and S.~L. Chin, \enquote{{Ultrafast white-light continuum generation
  and self-focusing in transparent condensed media},} J. Opt. Soc. Am. B
  \textbf{16}, 637 (1999).

\bibitem{Alfano2006}
R.~R. Alfano, ed., \emph{{The Supercontinuum Laser Source}} (Springer-Verlag,
  New York, 1989), 2nd ed.

\bibitem{Eichert2014}
M.~A.~R. Eichert, H.~H.~U. Onghua, M.~A. R.~F. Erdinandus, M.~A.~S. Eidel,
  P.~E. N. G.~Z. Hao, T.~R. R.~E. Nsley, D.~A.~P. Eceli, J.~E. M.~R. Eed, D.~M.
  A.~F. Ishman, S.~C.~W. Ebster, D.~A. J.~H. Agan, and E.~R. I. C. W. V. A.
  N.~S. Tryland, \enquote{{Temporal, spectral, and polarization dependence of
  the nonlinear optical response of carbon disulfide},} Optica \textbf{1},
  436--445 (2014).

\bibitem{patent}
International Standard ISO 12233, Photography - Electronic still-picture cameras - Resolution measurement, ISO, 2000.
  
\bibitem{Li2009}
T.~Li, H.~Feng, Z.~Xu, X.~Li, Z.~Cen, and Q.~Li, \enquote{{Comparison of
  different analytical edge spread function models for MTF calculation using
  curve-fitting},} Proc. SPIE \textbf{7498}, 74981H (2009).

\end{thebibliography}
\end{document}